\documentclass[conference]{IEEEtran}

\usepackage[letterpaper, left=1in, right=1in, bottom=1in, top=0.75in]{geometry}
\usepackage{booktabs} 
\usepackage[linesnumbered,ruled]{algorithm2e}
\usepackage{amsmath}
\usepackage{mathrsfs}
\usepackage{multirow}
\usepackage{subfig}
\usepackage{ragged2e}
\usepackage{cite}
\usepackage{graphicx} 
\usepackage{epstopdf}
\usepackage{epsfig}

\newcommand*{\permcomb}[4][0mu]{{{}^{#3}\mkern#1#2_{#4}}}
\newcommand*{\perm}[1][-3mu]{\permcomb[#1]{P}}

\begin{document}
%
\title{Room Recognition Using Discriminative Ensemble Learning with Hidden Markov Models for Smartphones}

\author{\IEEEauthorblockN{Jos\'e Luis Carrera V.}
\IEEEauthorblockA{Institute of Computer Sciences\\University of Bern\\
Switzerland\\
Email: carrera@inf.unibe.ch}
\and
\IEEEauthorblockN{Zhongliang Zhao}
\IEEEauthorblockA{Institute of Computer Sciences\\University of Bern\\
Switzerland\\
Email: zhao@inf.unibe.ch}
\and
\IEEEauthorblockN{Torsten Braun}
\IEEEauthorblockA{Institute of Computer Sciences\\University of Bern\\
Switzerland\\
Email: braun@inf.unibe.ch}
}
\maketitle
\begin{abstract}
An accurate room localization system is a powerful tool for providing location-based services. Considering that people spend most of their time indoors, indoor localization systems are becoming increasingly important in designing smart environments. In this work, we propose an efficient ensemble learning method to provide room level localization in smart buildings. Our proposed localization method achieves high room-level localization accuracy by combining Hidden Markov Models with simple discriminative learning methods. The localization algorithms are designed for a terminal-based system, which consists of commercial smartphones and Wi-Fi access points. We conduct experimental studies to evaluate our system in an office-like indoor environment. Experiment results show that our system can overcome traditional individual machine learning and ensemble learning approaches.
\end{abstract}
%
\IEEEpeerreviewmaketitle
\section{Introduction}
Due to the current importance of context aware services and the growing ubiquitousness of the Internet of Things (IoT), indoor localization has become an interesting research topic. Moreover, with the increase of smartphone devices, indoor localization applications for smartphones have attracted attention. Often, indoor context aware services require higher localization accuracy than outdoor services.
Additionally, the algorithmic complexity of a localization application is constrained by the limited computation and power resources on the smartphone. Thus, indoor localization is still considered as an open challenging problem. 
With the development of smartphones and the availability of more embedded sensors on mobile devices, several indoor localization methods (e.g, Wi-Fi or magnetic field-based fingerprinting, Wi-Fi or Bluetooth-based ranging, etc.) have been proposed. However, due to the widely extended availability of Wi-Fi signals in indoor environments, Wi-Fi radio-based localization has attracted most attention \cite{ARealTimeIndoorTracking}. Wi-Fi received signal strength indicator (RSSI) is the most used parameter for indoor localization. It can be used both in range-based or fingerprinting-based approaches.

Indoor environments provide many different ubiquitous radio signals, such as Wi-Fi, Bluetooth, magnetic field, sound, light, etc \cite{SemanticSLAM}. The earth magnetic field (MF) has distortions over space due to the presence of ferromagnetic materials. These MF distortion patterns can be also used to identify indoor locations \cite{SemanticSLAM}. Thereby, MF and Wi-Fi observations can be used as radio fingerprints to detect unique locations in indoor environments. 

Fingerprinting-based indoor localization systems usually consist of two phases: training phase (off-line) and localization phase (on-line). In the off-line phase, the fingerprint database is built by collecting various types of radio signals in the target indoor environments. In the on-line phase, the observed fingerprint at an unknown location is compared with the stored fingerprints in the fingerprint database to determine the closest match (i.e, prediction). In the on-line phase, any single discriminative learning model can be applied. However, ensemble learning models usually allow the production of better predictive performance compared to single models.  Fingerprinting-based methods that build the classification model exclusively based on beforehand observed data (i.e., fingerprint database) are named discriminative learning methods.  

\begin{figure}
	\centering
	\includegraphics[scale=0.29]{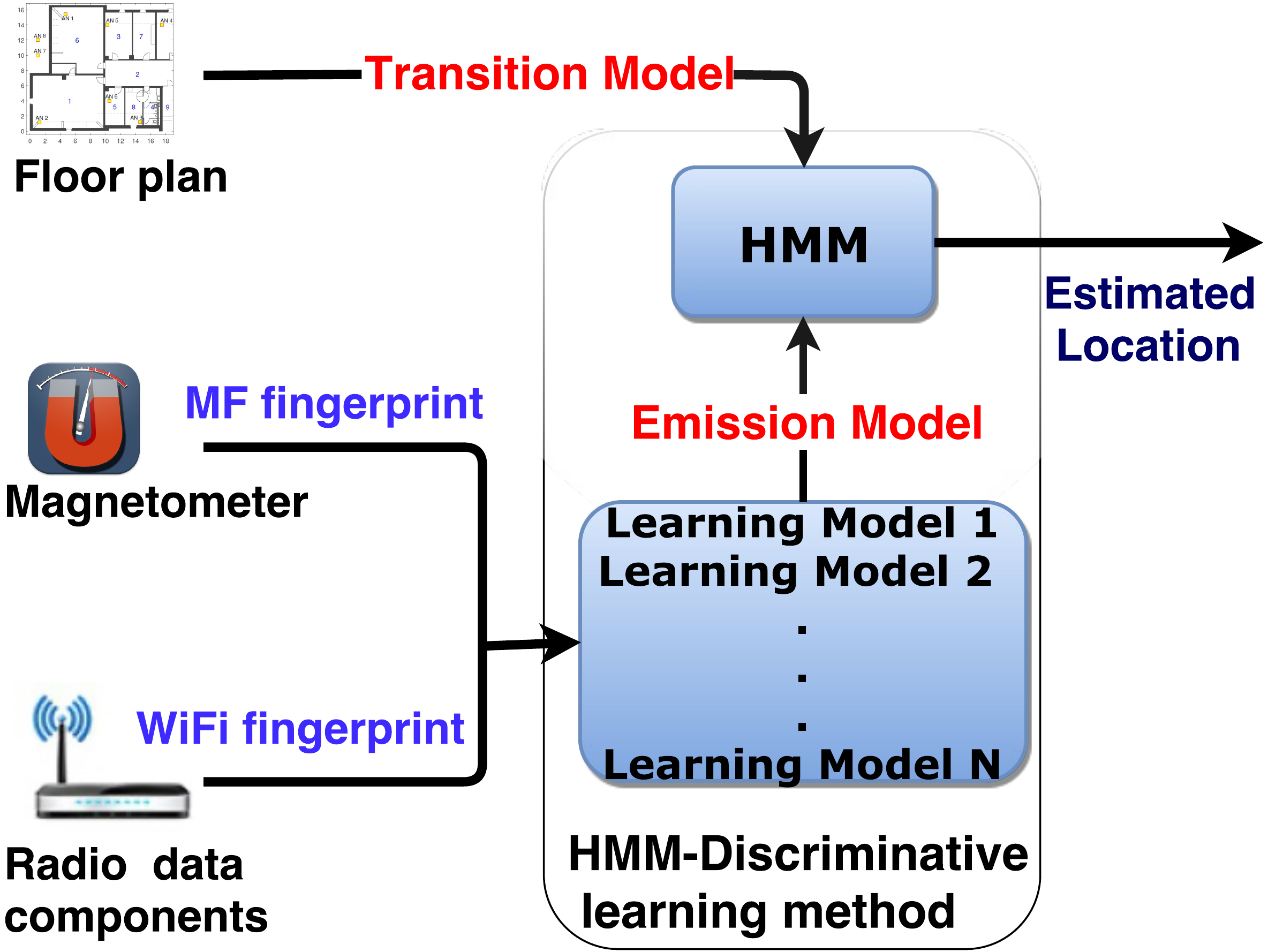}
	\caption{Indoor Localization System Architecture. Integration of HMM with individual learning models.}
	\label{Systemoverview}
\end{figure}

We present a novel room-level localization approach by fusing Wi-Fi RSSI, MF and coarse-grained floor plan information in an ensemble discriminative learning model. To achieve high and stable performance, we apply Hidden Markov Models (HMM) and discriminative learning models to integrate Wi-Fi, MF readings and information about transitions between rooms to achieve room level localization. Our approach requires only information about the physical distribution of rooms in the target area. Therefore, a precise floor plan is not needed. Thus,  unlike traditional fingerprinting localization approaches, we include room transition information in the localization process.  Figure \ref{Systemoverview} shows an overview of our proposed approach. The indoor localization system tracks room level location of a person holding a smartphone in real-time. It works as a basis for indoor location-based services for IoT. The main contributions of this work are summarized as follows.
\begin{itemize}
\item We propose a novel room level localization method based on an enhanced learning model. Our ensemble learning model  fuses Wi-Fi, MF and room transition information to achieve high room level localization accuracy. 
\item We combine a set of individual machine learning methods in an ensemble learning model. Our approach integrates HMM with discriminative learning techniques to achieve high prediction performance. Unlike traditional Wi-Fi fingerprint-based approaches using individual machine learning methods, we include room transition information in the localization process. 
\item We implement an efficient terminal-based indoor localization system for smartphones that is able accurately track in real-time room level location of a person holding a smartphone. Since our approach provides room level localization, the fingerprint database is built by taking Wi-Fi and MF measurements while walking randomly through the environment, which requires only room-labeled samples that can be collected in a very short time period. Thus, the off-line phase becomes a simple process.  
\item  We perform a set of experiments to validate the performance of our localization method. We evaluate our system in a complex environment along three different moving paths. 
\end{itemize}
The rest of the paper is organized as follows. In Section II we present some related work. The ensemble learning model for indoor localization is reviewed in Section III. Section IV presents the implementation of the terminal-based system. Section V presents the performance evaluation results of our approach. Section VI concludes the paper.

\section{Related Work}

Pedestrian Dead Reckoning (PDR) has attracted research interest due to the fast development of Inertial Measurement Units (IMUs) in modern smartphones. IMUs can be leveraged to implement pedestrian movement detection such as heading orientation, step recognition, stride length estimation \cite{PDR1}. Thus, by using IMUs, PDR systems estimate the new location based on the previously determined location. For instance in ~\cite{PDR2,PDR3}, the heading orientation is determined based on gyroscope measurements, whereas the displacement is estimated from accelerometer readings. In \cite{PDR2}, a Heuristic Drift Elimination (HDE) is intended to deal with accumulated errors. However, HDE requires specialized sensors deployed on the foot of the pedestrian.  In \cite{PDR4}, authors use readings of the gyroscope to recognize physical turns of the pedestrian, whereas the stride length is determined by readings of the accelerometer. PDR systems measure position changes rather than the absolute position, which results in an accumulation of sensor errors over time. Therefore, these systems must consider additional information to deal with this type of errors.


Radio signals are often used to provide indoor localization. Several parameters of radio signals can be leveraged to locate targets. In \cite{RDS1} for instance, the authors propose to use received signal strength indicator (RSSI), whereas in \cite{RDS2} time information related to radio propagation is applied.
Radio-based indoor localization can be classified as range-based and range-free methods. Range is defined as the propagation distance from the target to Anchor Nodes (AN). The first stage in range-based localization methods  is to calculate the propagation distances, which is called ranging.  Then,
different positioning algorithms can be used to estimate the absolute locations of the targets, such as trilateration and multilateration \cite{RDS3}. However, unlike outdoor localization, trilateration does not work well in indoor environments because of the presence of obstacles and room partitions\cite{RDS4}. In range-free localization methods, fingerprinting \cite{RDS5} is very often used because of its robustness to multipath propagation. However, it is very time consuming to build up a radio map, which is required to locate the targets in fingerprinting. In \cite{WiFiEnsemble}, a Wi-Fi-based ensemble learning model is proposed. Authors proposed to provide localization by developing a room-based weighted method in an ensemble learning technique. The room detection method relies on a simple average of the coordinates output by a k-NN estimator. Therefore, this process is prone to errors. Typically, ranging requires much less labor efforts than building radio maps for fingerprinting approaches.

Applications of HMM can be considered to provide  indoor localization. In \cite{HMM1}, authors employed HMM and radio propagation models to reduce the calibration efforts. The system utilizes a discrete probability distribution to derive probable positions. Then, the position is estimated from the most probable estimated positions. In \cite{HMM2}, authors include movement measurements (e.g., heading orientation) in the proposed HMM. Thus, the reported accuracy is improved compared to \cite{HMM1}. In \cite{HMM3}, authors propose to fuse IMU measurements with wireless signal readings. Then, the candidate position is derived based on the pedestrian's motion pattern and the most probable  wireless signals at that position. Despite authors report good accuracy, the method to determine the transition probabilities is not explained. Moreover, the applicability of the solution is restricted to the accuracy of the pedestrian motion detection method. In \cite{HMM4}, authors propose to fuse a RSSI pattern recognition method and HMM to provide indoor localization. The pattern recognition method uses RSSI variation instead of raw RSSI. Then, the transition probabilities of the HMM are derived from the pedestrian trajectories and the pattern recognition method. The pattern recognition method relies on a beforehand built radio map database. Thus, some reference locations are defined through the indoor environment in an off-line phase to collect reference samples. Such collection process could take severals hours or days for small or big areas, which is very labor expensive and time consuming.

\section{HMM-discriminative Ensemble Learning Method} \label{HMM-ensemble}
\begin{figure}
	\centering
	\includegraphics[scale=0.28]{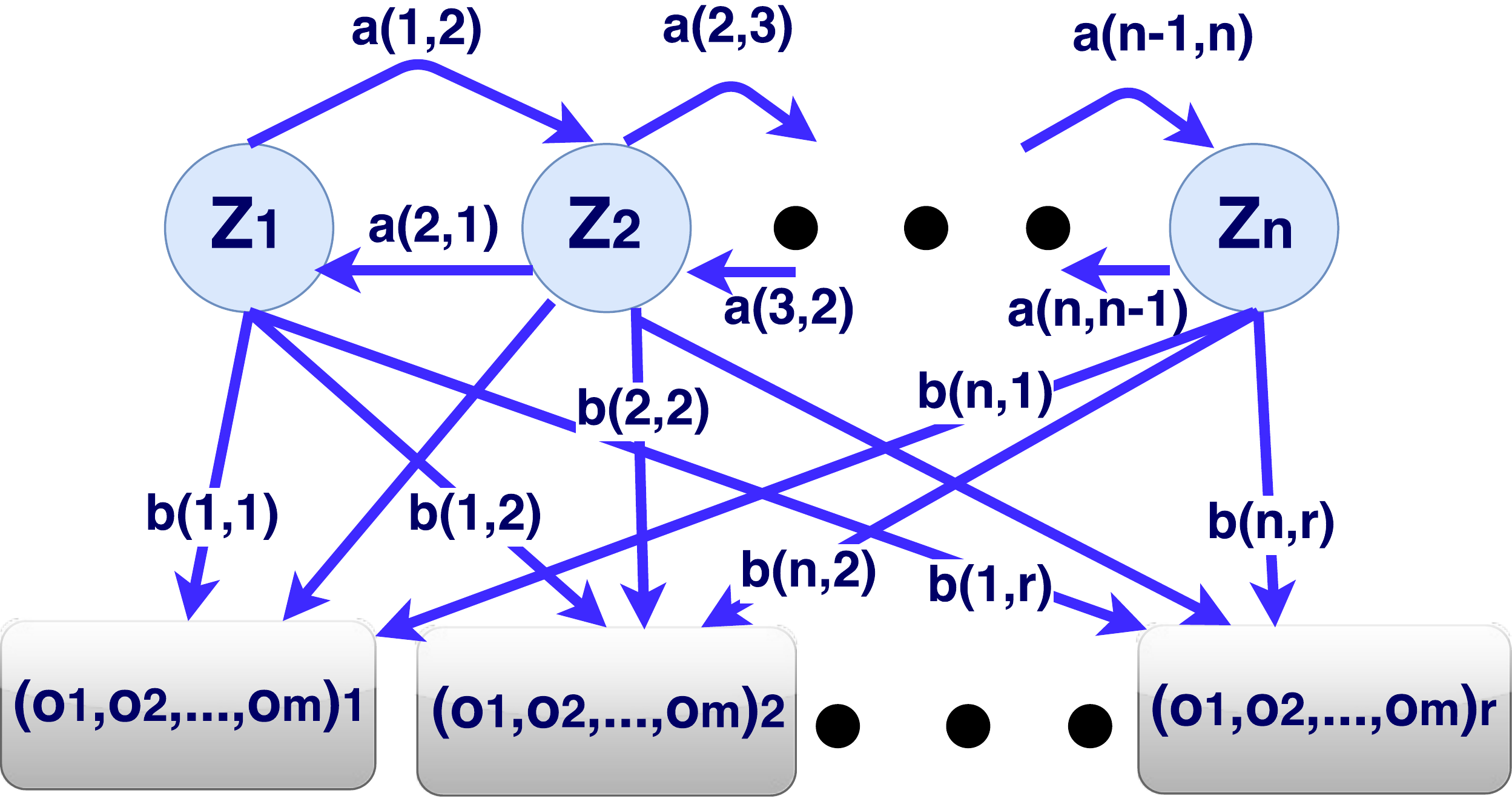}
	\caption{Probabilistic parameters of the proposed Hidden Markov Model.}
	\label{HMMparameters}
\end{figure}
Ensemble learning methods intend to improve prediction results by combining  several individual machine learning techniques into one predictive model. Thus, ensemble approaches usually achieve better predictive performance compared to single models. We propose an ensemble learning model by combining HMM with discriminative machine learning methods to provide accurate room detection for smartphone users. Figure \ref{HMMparameters} shows the probabilistic parameters of the proposed HMM. It is worth to notice that this model is suitable for any zone detection (i.e., any subarea in the area of interest). Hereafter, we will refer to room as zone. The proposed HMM is based on the concept of Markov localization \cite{MC-LOCALIZATION}, which can be described as estimating the state of HMM with controllable state transitions. For the localization problem, we refer to zones as states of the HMM. Thus, our HMM is specified by the following components:
\begin{itemize}
 \justifying
 \item A set of \textit{n} states $Z=\{z_1,z_2,...,z_n\}$, where $z_i$ is the identifier value of the zone $i$. Thus, the discrete random variable $x_t \in Z$ represents the hidden state at time $t$.
 
\item A transition probability matrix $A$,
\begin{gather}
A=\begin{pmatrix} 
	a_{1,1}&a_{1,2}&...&a_{1,n}\\
    a_{2,1}&a_{2,2}&...&a_{2,n}\\
	. & . & ...&.\\
	. & .&...& .\\
    . & .& ... & .\\
    a_{n,1}&a_{n,2}&...&a_{n,n}\\
\end{pmatrix},
\end{gather}
where $a_{ij}$ represents the probability of moving from zone $z_i$ to zone $z_j$. 
        
    \item A set of observations $O$,
    \begin{equation}
    O=\{(o_1,o_2,...o_m)_1,...(o_1,o_2...o_m)_r\},
    \end{equation}
where $o_i$ is the zone prediction result of the $i$-th individual machine learning method. Therefore, $O$ is a set of $\perm{r}{m}$ permutations with repetitions allowed, where $r$ is the number of zones and $m$ is the number of individual machine learning methods used to build the ensemble predictive model. Conditional independence must be assured among the individual machine learning methods. Thus, the random variable $y_t\in O$ represents the observations at time $t$. 
    
    \item An emission probability matrix $B$ of observation likelihoods, which is also called emission probabilities, 
\begin{gather}   
 B=\begin{pmatrix} 
	b_{1,1}&b_{1,2}&...&b_{1,r}\\
    b_{2,1}&b_{2,2}&...&b_{2,r}\\
	. & . & ...&.\\
	. & .&.& .\\
    . & .& ... & .\\
    b_{n,1}&b_{n,2}&...&b_{n,r}\\
\end{pmatrix},
\end{gather}
where $b_{i,j}$ represents the probability of an observation $(o_1,o_2...o_m)_j$ being generated at zone $z_i$.

    \item An initial probability distribution $\pi=\pi_1,\pi_2,...,\pi_n$ over zones.

\end{itemize}
The individual learning methods only rely on the latest observed fingerprint of Wi-Fi RSS and MF readings for localization, which may produce incorrect prediction results. However, the HMM can be used to integrate zone transition information and the current observed information (e.g., current observed fingerprint) to improve prediction results. Hereafter, we will refer to our HMM-discriminative learning model as HMM-d model.
In any model with hidden variables (e.g., HMM), the task of determining the sequence of variables (e.g., zones) that is the underlying source of some sequence of observations is named the decoding task. Thus, given a sequence of observations $y_{t-i},...,y_{t-1},y_{t}$, and a settle model HMM $\lambda=\{\pi,A,B\}$, the sequence of hidden states $x_{t-i},...,x_{t-1},x_{t}$ can be estimated by employing the Viterbi algorithm \cite{viterbi}.

\subsection{Transition Probabilities}
The transition probabilities express the likelihood of moving from one state (i.e., zone) to another. Zones must be defined beforehand. Connections among  zones in the coarse-grained floor plan determine the transition probabilities. Therefore, the transition probability matrix can be written as follows:

\begin{equation}
A=\{a_{ij}=P(x_{t+1}=z_j \mid x_{t}=z_i)\},
\end{equation}
where $A$ is a $n\times n$ matrix, $a_{ij}$ represents the transition likelihood between zone $z_i$ to zone $z_j$. Therefore, $\sum_{j=1}^n a_{ij}=1$.  

\subsection{Emission Probabilities}
The emission probability is the likelihood of producing a particular set of observations $y_j$ at zone $z_i$. Therefore, the  emission probability matrix can be written as follows:

\begin{equation}
B=\{b_{ij}=P(y_j \mid z_i )\}, \forall y_j \in O \land z_i \in Z,
\end{equation}
where $y_j=(o_1,o_2...o_m)_j$ and $o_i$ is the zone prediction result from the $i$-th individual learning method. Since individual machine learning methods are different and independent of each other, it is reasonable to assume that their outcomes are conditionally independent. Therefore, $b_{ij}$ can be written as follows:

\begin{equation}
b_{ij}=\prod_{n=1}^n P(o_{j}\mid z_i)_n, 
\end{equation}
where $P(o_{j}\mid z_i)_n$ is the probability of predicting $o_{j}$ at zone $z_i$ by the $n$-th individual discriminative learning method. Therefore, $P(o_{j}\mid z_i)_n$ represents the sensitivity of the individual learning method $n$ at zone $i$. Thus, $P(o_{j}\mid z_i)_n$ can be written as follows: 

\begin{equation}
P(o_{j}\mid z_i)_n=\frac{TP_n}{TP_n+FN_n}, 
\end{equation}
where $TP_n$ and $FN_n$ are the true positive and false negative rate of the $n$-th individual discriminative learning method.

\begin{figure*}[h]
\centering
\null\hfill
	\subfloat[Zone definition and graphical transition model]
   {\includegraphics[scale=0.18]{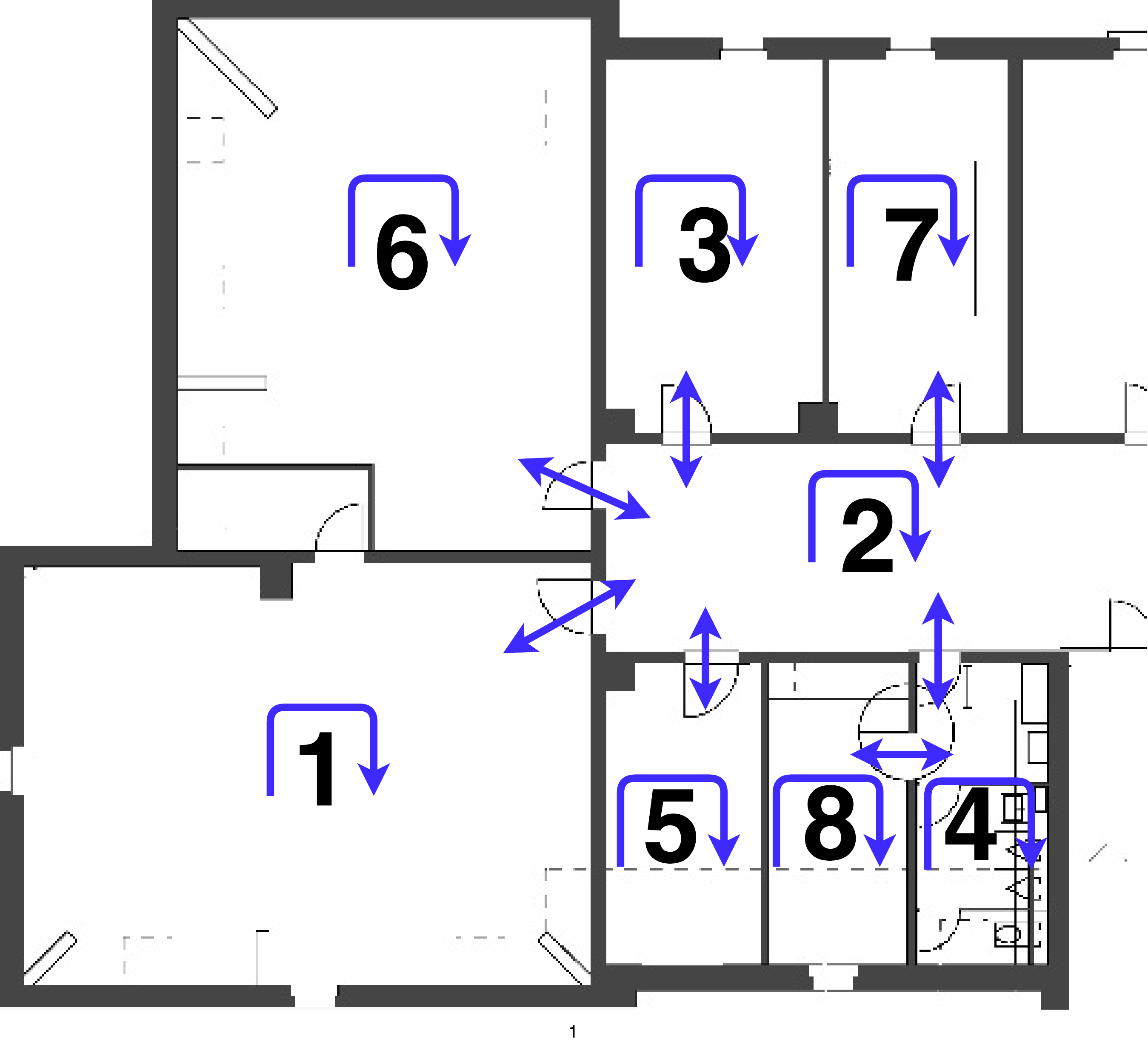}\label{HMM}}
    \hfill
    \subfloat[Transition model for HMM-d]
    {\includegraphics[scale=0.19]{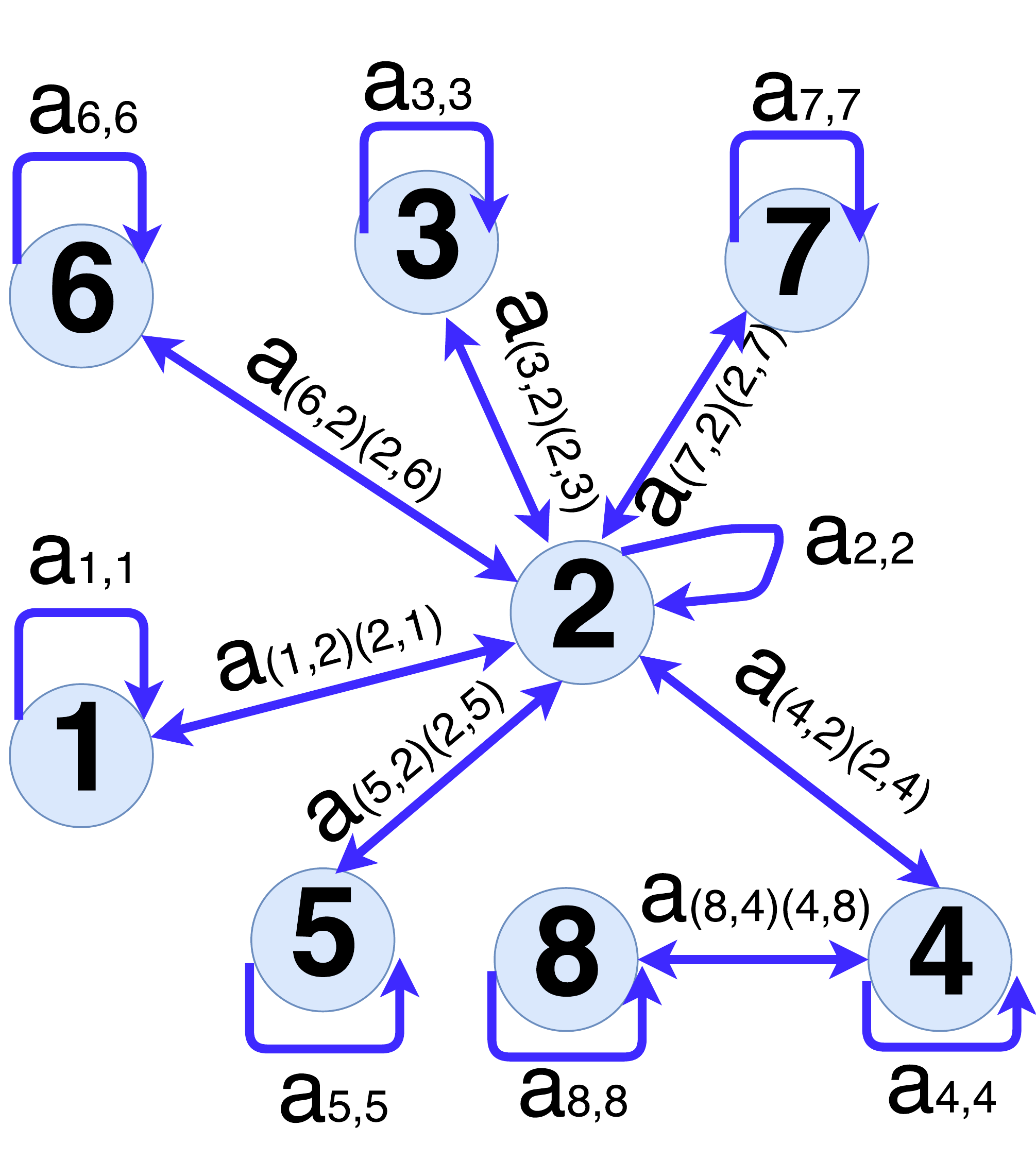}\label{HMM1}}
	\hfill\null
	\caption{Zone definition and transition model for HMM-d.}
	\label{transitionModel}
\end{figure*}

\section{Implementation}
We have implemented a terminal-based system for accurate indoor localization. The system comprises two main components: a mobile target and many Wi-Fi Anchor Nodes (ANs). The proposed localization algorithms are running on the mobile target. Figure \ref{Systemoverview} presents the overview of the system.
ANs are some commercial Wi-Fi access points deployed at known or unknown locations along the area of interest. To provide the maximum coverage inside the area of interest, the locations of ANs are defined by the boundary corners and the boundary itself. We have adopted D-Link D-635 and D-Link DAP-2553 devices as Wi-Fi ANs in this work. The beacon period is configured to $100ms$ for ANs. 
The mobile target can be any commercial Android smartphone, which supports Wi-Fi RSSI readings and magnetic field sensor readings. We have deployed the localization algorithms in a Motorola Nexus 6 smartphone. Hereafter, we refer to  Motorola Nexus 6 as the mobile target (MT). In order to save resources in the smartphone, we set the sampling rate of the magnetic field sensor to 14Hz. However, the Wi-Fi sampling frequency is much lower, i.e., 3Hz. Table \ref{table_smarphone} shows the main characteristics of the mobile target used in this work.
Additionally, it is necessary to have coarse-grained information about the area of interest. The system requires information related with  zone distribution and physical connections among zones (i.e., zone transition information). The system reports the location of the target in real time. Zone information is also included in the coarse-grained floor plan (i.e., how the area of interest is split in zones). Figure \ref{transitionModel} shows the zone definition and transition model in our system. The basic assumption to compute matrix $A$ is that the likelihood of staying at the same zone is higher than the likelihood of going to another one. Thus, the transition probability matrix $A$ was empirically defined as follows: 
\begin{gather*}
A=\begin{pmatrix} 
	0.6&0.4& 0& 0& 0& 0& 0& 0\\
	0.1 & 0.4 & 0.1& 0.1& 0.1& 0.1& 0.1& 0\\
	0 & 0.4 & 0.6& 0& 0& 0& 0& 0\\
	0 & 0.2 & 0& 0.6& 0& 0& 0& 0.2\\
    0 & 0.4 & 0& 0& 0.6& 0& 0& 0\\
    0 & 0.4 & 0& 0& 0& 0.6& 0& 0\\
   0 & 0.4 & 0& 0& 0& 0& 0.6& 0\\
    0 & 0 & 0& 0.4& 0& 0& 0& 0.6\\ 
\end{pmatrix},
\end{gather*}
\begin{table}
	\centering
	\caption{Mobile Target Specifications}
	\begin{tabular}{|l|} \hline
	\textbf{Specification}\\ \hline
	\textbf{Model:}  Motorola Nexus 6; \textbf{OS:} Android 5.1.1\\
	\textbf{CPU:} Quad-core 2.7 GHz; \textbf{RAM:} 3GB\\
	\textbf{WLAN:}Wi-Fi a/b/g/n
	\textbf{Magnetometer:} Res.: 0.15 Ran.: 9830\\
	   \hline\end{tabular}
	\label{table_smarphone}
\end{table}

To ensure conditional independence between the individual learning methods in HMM-d, we setup three completely different discriminative machine learning techniques for the zone prediction method. KStar  \cite{KSTAR}, Multilayer Perceptron (MLP) \cite{MLP} and the J48 decision tree \cite{J48} machine learning algorithms were selected. Moreover, each individual machine learning method was trained with independent balanced training dataset, i.e., same number of instances for every class (zone). Since HMM-d is a zone level (i.e., room recognition) detection method, it is not needed to define any fixed survey point to build the fingerprint database. Thus, the fingerprint database for zone prediction is built by taking Wi-Fi and magnetic field measurements while walking randomly through the environment, which requires only zone labeled samples in a very short time period. Thus, the off-line phase becomes a simple process.  

Internal parameters of learning-based algorithms are optimized from training data. Additionally, certain algorithms also have parameters that are not optimized during the training process. These parameters are called hyperparameters of the learning-based algorithm. Since hyperparameters have significant impact on the performance of the learning-based algorithm, we use a nested cross validation technique to adjust them \cite{MLTechniques}. Nested cross validation techniques define an inner and outer cross validation. The inner cross validation is intended to select the model with optimized hyperparameters, whereas outer cross validation is used to obtain an estimation of the generalization error. Ten-fold cross validation was applied on both inner and outer cross validation. The classifiers were optimized over a key of hyperparameters. We optimized the global blend percentage ratio hyperparameter for KStar \cite{KSTAR}, the confidence factor for J48 \cite{J48}, as well as number of hidden layers and neurons per layer for MLP \cite{MLP}. Based on the parameter optimization process, we established the optimal hyperparameter values for the classifiers as follows: global percent ratio of 30\% for KStar, single hidden layer with 10 neurons for MLP, and confidence factor of 0.25 for J48.

\section{Performance Evaluation}
\subsection{Measurement Setup}
\begin{figure*}[htp]
	\centering
	\subfloat[The First Trajectory \label{FirstTrajectory}]{\includegraphics[scale=0.13]{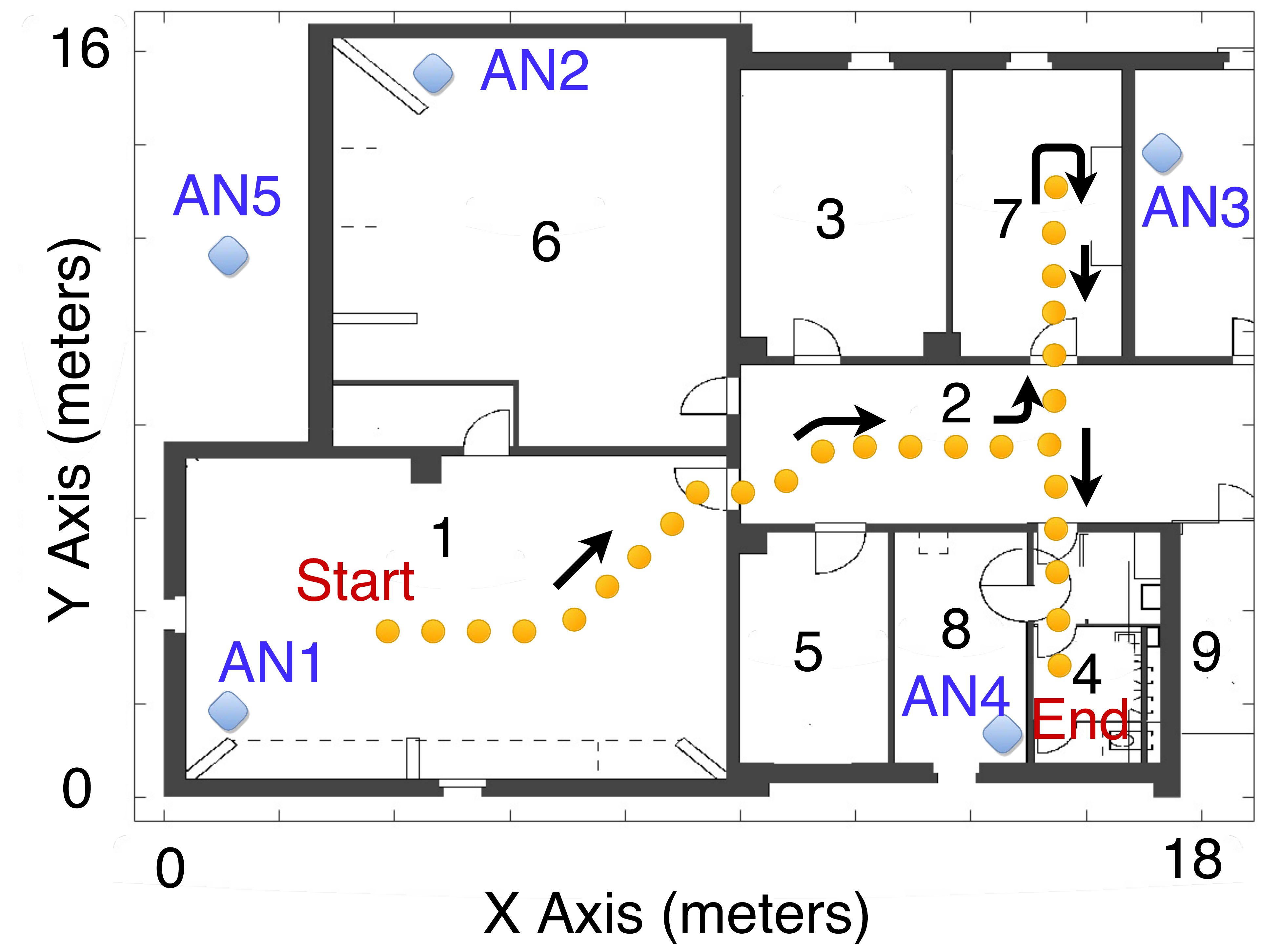}}
	\subfloat[The Second Trajectory]{\includegraphics[scale=0.13]{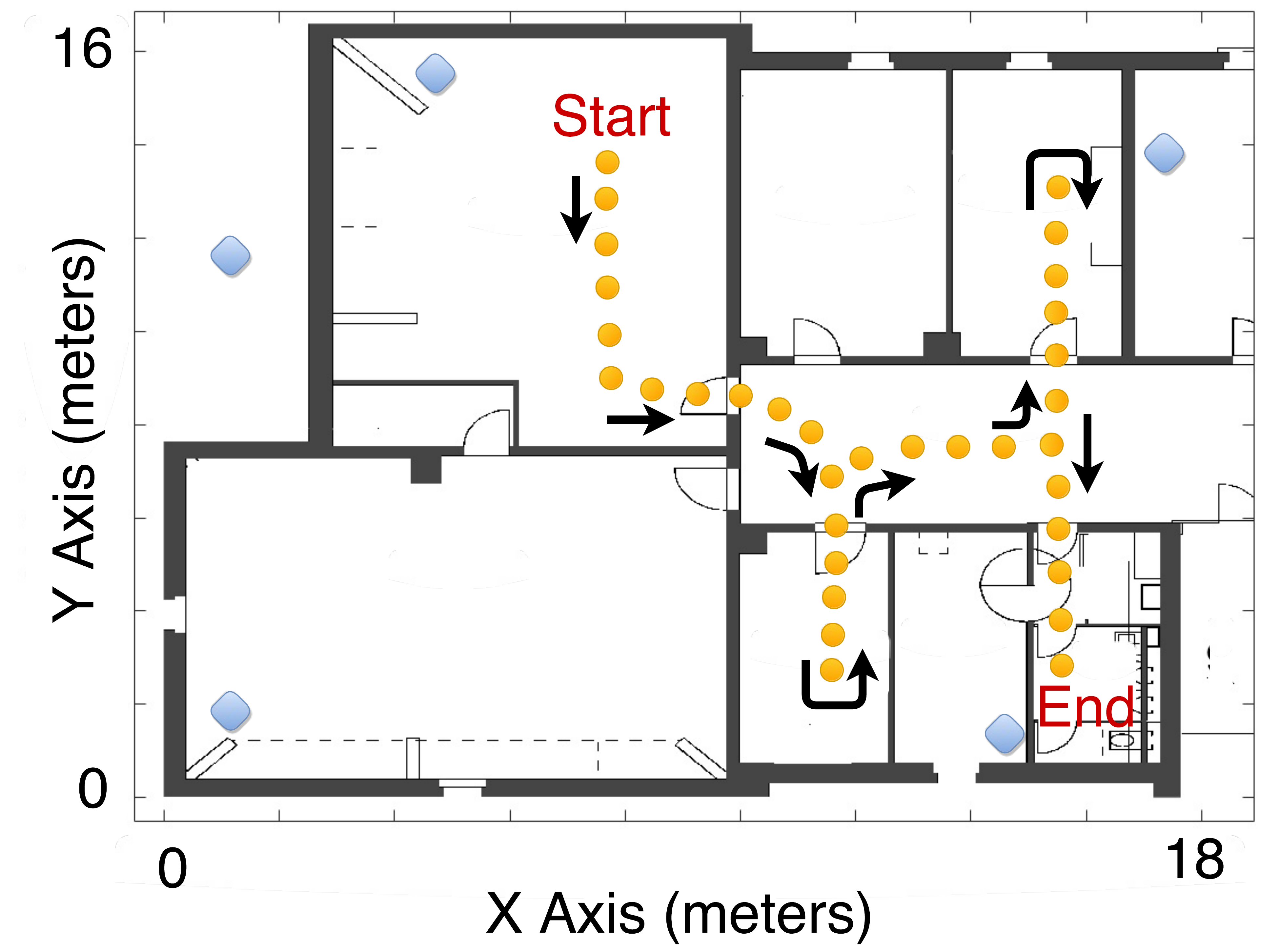}}
    \subfloat[The Third Trajectory]{\includegraphics[scale=0.13]{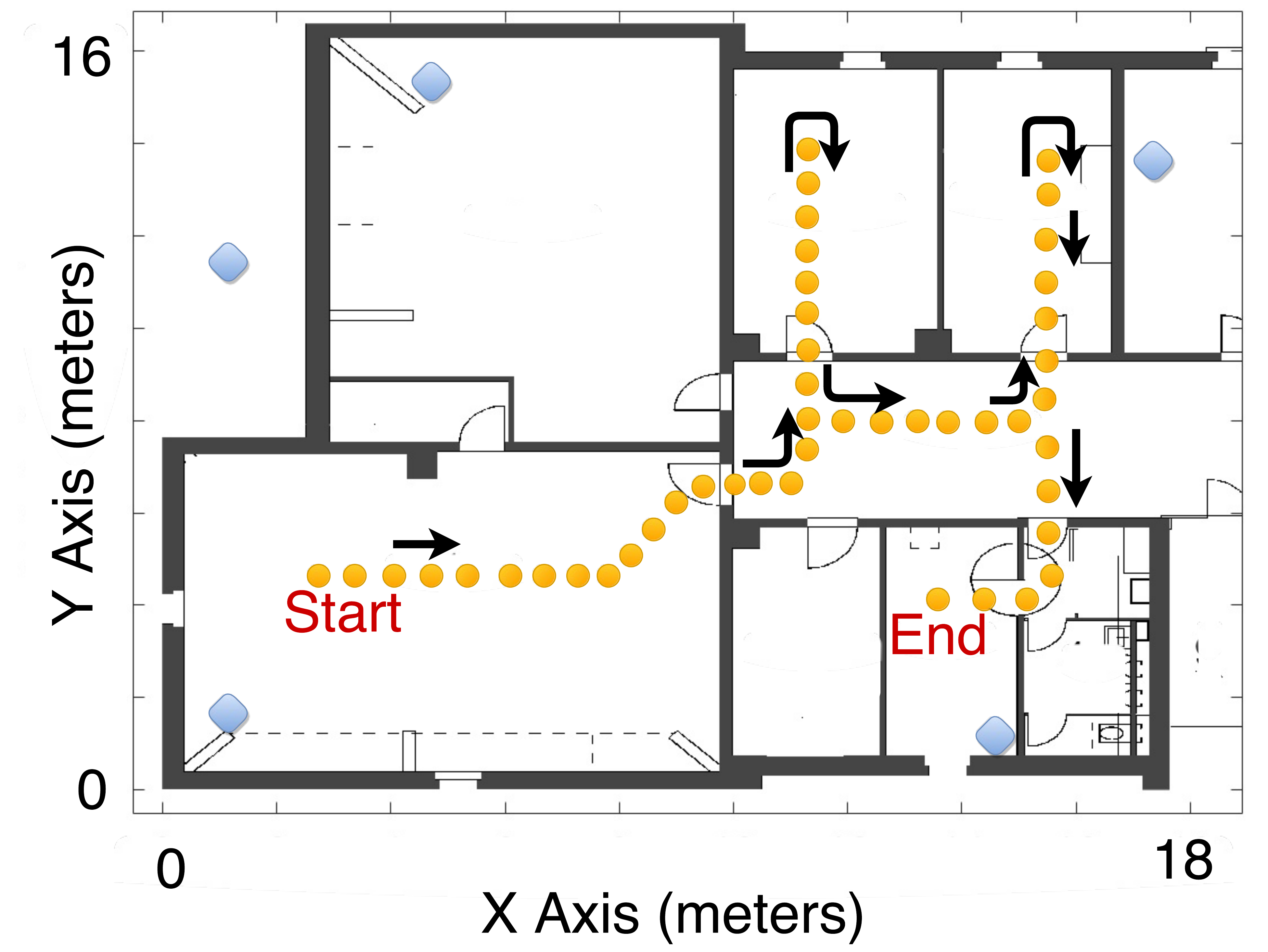}}
	\caption{Zones, trajectories and ANs distribution (Diamond points: Anchor Nodes; Yellow points: trajectories)}
	\label{scenariosAnDistribtion}
\end{figure*}  
Experiments were conducted in the building of the Institute of Computer Science at the University of Bern. A part of the third floor with an area of $288 m^2$ ($18 m$ x $16 m$) was chosen to deploy the  localization system. The smartphone is held by a person moving along three different trajectories (Figure \ref{scenariosAnDistribtion}). The zone detection method is launched every time a new fingerprint measurement is available (i.e., approximately twice per second). We define 8 zones in our environment. Each zone is a wall separated area (i.e., rooms, corridor).  Figure \ref{scenariosAnDistribtion} presents the physical distribution of zones, ANs, and trajectories. Additionally, to compare HMM-d to another ensemble learning model, we implemented a majority voting-based method. Hereafter, we refer to majority voting-based method as Voting method. The Voting method uses predicted zone labels from KStar, J48, and MLP for the majority voting rule. Further details about majority voting-based methods can be found in \cite{VotingMethod}. All the algorithms use the same fingerprint database of Wi-Fi RSS and MF readings, which have been measured during the data collection procedure. The individual predictors can be regarded as traditional fingerprint-based approaches, while the proposed HMM-d is a new ensemble predictor.

\subsection{Room Level Localization Accuracy}
\begin{figure}
	\centering
	\includegraphics[scale=0.28]{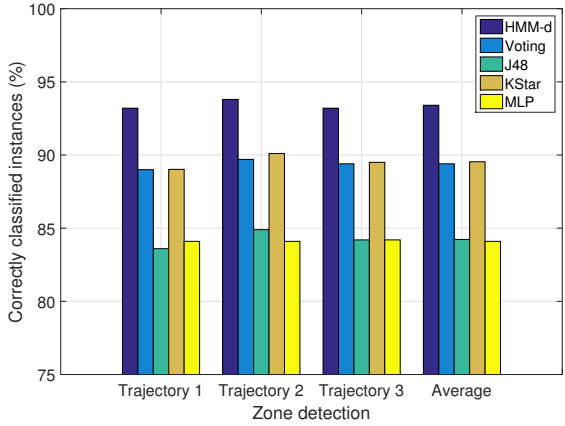}
     \vspace{-12pt}
	\caption{Predictive Model Accuracy}
	\label{ZoneDetectionAccuracy}
\end{figure}

\begin{figure}
	\centering
	\includegraphics[scale=0.27]{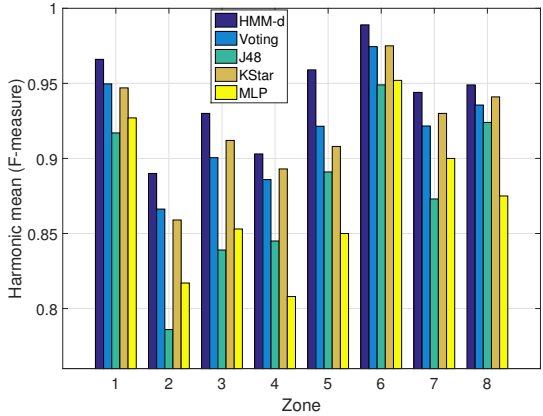}
	\caption{Zone Detection Performance F1 score }
	\label{FMeasureZoneDetection}
\end{figure}

\begin{figure}
	\centering
	\includegraphics[scale=0.30]{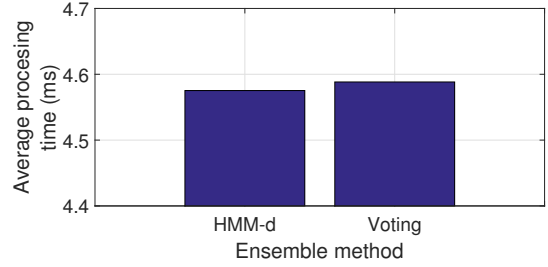}
	\caption{Average processing time (per prediction) of ensemble methods.}
	\label{procesingTime}
\end{figure}
To evaluate our prediction models, we consider three  measures: prediction accuracy, F1 score, and processing time. In classification problems, accuracy is the ratio of correctly predicted observation to the total observations. F1 is the harmonic mean of precision and sensitivity. Precision is defined as the number of true positives (TP) divided by TP and the number of  false positives (FP): TP/(TP+FP).  Sensitivity is defined as the number of TP divided by TP and the number of  false negatives (FN): TP/(TP+FN) \cite{accuracyMeasures}. Thus, F1 considers both performance measures, precision, and sensitivity. F1 can be written as follows: 

\begin{equation}
F1=2 \cdot \frac{sensitivity \cdot precision}{sensitivity+precision}, 
\end{equation}

Figure \ref{ZoneDetectionAccuracy} shows the accuracy of zone prediction for the five predictors in the three trajectories. Figure \ref{FMeasureZoneDetection}  presents the F1 score by zone for the learning algorithms. Figure \ref{procesingTime} shows the average of prediction processing time for both ensemble models HMM-d and Voting. Due to the hyperparameter optimization process, performance accuracy of the individual learning models (i.e., KStar, J48 and MLP) is higher than $80\%$. However, results show a clear improvement between the individual learning models and our ensemble learning model HMM-d. As it can be seen in Figure \ref{ZoneDetectionAccuracy}, our proposed HMM-d model outperforms KStar, J48, and MLP algorithms in the three tested trajectories. Accuracy of HMM-d is improved by $9.17\%$, $4\%$, $9.3\%$, and $9.2\%$ compared to J48, KStar, MLP, and the Voting method respectively. Unlike Voting, HMM-d is able to combine zone transition information with individual learning methods to improve prediction accuracy. 

Considering the F1 score,  HMM-d outperforms others in all tested zones (i.e., classes).  Therefore, HMM-d outperforms Voting, KStar, J48, and MLP considering accuracy and robustness. As it can be seen in Figure \ref{ZoneDetectionAccuracy}, accuracy of the learning algorithms remains similar (i.e., not significant difference) in all tested trajectories. However, if we consider sensitivity and precision as performance measure, it is clear to notice that zone 2 is the hardest zone to classify correctly  (as shown in Figure \ref{ZoneDetectionAccuracy}). This result is explained as zone 2 corresponds to the corridor (see Figure \ref{FirstTrajectory}). Thus, fingerprints obtained in this zone are very similar to fingerprints obtained in adjacent zones, especially in the border areas. However, HMM-d improves the classification accuracy in this zone by $10.2\%$,  $10.4\%$, $4\%$ and $8\%$ compared to Voting, J48, KStar, and MLP respectively.  Thus, by combining HMM, KStar, J48, and MLP, our  approach allows the production of better predictive performance compared to individual and ensemble voting models. Unlike Voting, the HMM-d model is able to achieve better prediction performance than individual learning methods in all zones. As shown in Figure \ref{procesingTime}, the average time of prediction is very similar in both HMM-d and Voting ensemble methods. However, HMM-d reduces the prediction processing time by $0.0125 ms$  compared to the Voting method. This means that considering transition zone information in the prediction process takes lower computational efforts than processing the voting rule \cite{VotingMethod} that is used in voting methods. Thus, HMM-d model overcomes Voting, KStar, J48, and MLP methods by accuracy, robustness an processing time.

\section{Conclusions}
In this work, we proposed a terminal-based room level indoor localization system, which integrates discriminative learning techniques in an ensemble learning method using ubiquitous Wi-Fi and magnetic field fingerprints. Our proposed ensemble learning model achieves high prediction performance by combining less accurate individual discriminative learning models. We define rooms as zones, and adopted the Hidden Markov Model to integrate information about transition probabilities between zones and discriminative learning methods. Thus, this work presents a probabilistic-based system to achieve high room level localization accuracy of a smartphone user,  who is moving in a multi room indoor environment.  We evaluated our system in a complex real-world indoor environment. Evaluation results indicate that our proposed approach is more accurate and robust than individual learning and majority voting-based models. 

\section*{Acknowledgment}
This work was partly supported by the Swiss National Science Foundation, project no. 154458.



%

\bibliographystyle{IEEEtranS}
\bibliography{bib}




\end{document}